\documentclass[12pt,thmsa,notitlepage]{article}
\usepackage{amsmath}
\usepackage{amssymb}
\usepackage{amsfonts}
\usepackage{sw20lart}
\usepackage[onehalfspacing]{setspace}
\usepackage{graphicx}

\input{tcilatex}

\begin{document}

\title{Extra-Dimensional Approach to Option Pricing and Stochastic Volatility%
}
\author{{\large M. Q. Truong\thanks{%
mtruong@fontbonne.edu}} \\
{\small Department of Biological and Physical Sciences }\\
{\small Fontbonne University}\\
St. Louis, MO 63105}
\maketitle

\begin{abstract}
The generalized $5D$ Black-Scholes differential equation with stochastic
volatility is derived. The projections of the stochastic evolutions
associated with the random variables from an enlarged space or superspace
onto an ordinary space can be achieved via higher-dimensional operators. The
stochastic nature of \ the securities and volatility associated with the $3D$
Merton-Garman equation can then be interpreted as the effects of the extra
dimensions. We showed that the Merton-Garman equation is the first excited
state, i.e. $n=m=1,$ within a family which contain an infinite numbers of
Merton-Garman-like equations.

\newpage
\end{abstract}

\section{\protect\bigskip Introduction}

The time-evolution of the option pricing has been well-studied starting with
the work of Black and Scholes\cite{Black-Scholes}. This pioneering result of
Black and Scholes was then generalized to include stochastic volatility by
Merton \cite{Merton} and Garman\cite{Garman}.

The methodology of high energy physics have been used in the analysis of the
option pricing problem \cite{Bouchaud}. The analysis of the problem of \
option pricing by the methods of theoretical physics provides additional
computational power to the field of mathematical finance. The Black-Scholes
partial differential equation and its generalization have been reinterpreted
under quantum mechanical formalism by \cite{Baaquie} and \cite{Linetsky},
where the Hamilton for the Merton-Garman equation was derived. The dynamics
of the option price of a security derivative can then be given in terms of
path integrals. This paper used methods of theoretical physics,
specifically, extra-dimmensional formalism \cite{Truong}, as applied to the
analysis of problems associated with option pricing.

The general outline of this paper is divided as follows. In section 2, the
generalized Black-Scholes equation is derived from the Langevin stochastic
differential equations. In section 3, we review extra-dimensional theories
and their applications. In section 4, we derive the Merton-Garman equation
via extra-dimensional approach. And in section 5, conclusions and future
outlook are drawn.

\section{Derivative with Stochastic Volatility}

A security is any financial instrument that can be traded on the markets. A
security derivative is also a financial instrument that can be derived from
an underlying security and can also be traded on the markets. Some of these
security derivatives are futures, forwards and options. Suppose the value of
the option $f$ at time $t$ is given 
\begin{equation}
f=f(t,S(t),V(t),K,T),
\end{equation}%
where $S(t)$ is the value of the security at $t$, $V(t)$ is the variance of
the stochastic volatility, $K$ is the strike price, and $T$ is the time of
maturity. At time $t=T,$ the value of the option can be characterized by $%
f(T,S(T))$. The stochastic nature of the security and the volatility are
governed by the following coupled stochastic Langevin equations 
\begin{eqnarray}
\frac{dS(t)}{dt} &=&\phi S(t)+\sigma (t)S(t)R(t),  \label{langevin 1} \\
\frac{dV(t)}{dt} &=&\mu V(t)+\xi V(t)Q(t),  \label{langevin 2}
\end{eqnarray}%
where $\phi $ and $\mu $ are the drift rates associated with the security $%
S(t)$ and $V(t)$ respectively, $\sigma (t)$ is the stochastic volatility, $%
V(t)=\sigma (t)^{2}$ is the variance. The terms $R(t)$ and $Q(t)$ are the
correlated Gaussian white noise terms with zero means and the following
relations%
\begin{eqnarray}
\left\langle R(t)R(t^{\prime })\right\rangle &=&\left\langle Q(t)Q(t^{\prime
})\right\rangle =\frac{1}{\epsilon }\delta (t-t^{\prime }),  \label{RR noise}
\\
\left\langle Q(t)R(t^{\prime })\right\rangle &=&\left\langle R(t)Q(t^{\prime
})\right\rangle =\frac{\rho }{\epsilon }\delta (t-t^{\prime }),
\label{RQ noise}
\end{eqnarray}%
where $-1\leq \rho \leq 1$ is the correlation coefficient between $S(t)$ and 
$V(t).$ We now write down the second-order Taylor series expansion for the
total time derivative of the value of the option $f.$ The series expansion
yields%
\begin{eqnarray}
\frac{df}{dt} &=&\lim_{\epsilon \rightarrow 0}\frac{1}{\epsilon }\left\{
f(t+\epsilon ,S(t+\epsilon ),V(t+\epsilon ))-f(t,S(t),V(t))\right\} \\
\frac{df}{dt} &=&\frac{\partial f}{\partial t}+\frac{\partial f}{\partial S}%
\frac{dS}{dt}+\frac{\partial f}{\partial V}\frac{dV}{dt}  \notag \\
&&+\lim_{\epsilon \rightarrow 0}\frac{\epsilon }{2!}\left\{ \frac{\partial
^{2}f}{\partial S^{2}}\left( \frac{dS}{dt}\right) ^{2}+\frac{\partial ^{2}f}{%
\partial V^{2}}\left( \frac{dV}{dt}\right) ^{2}+\frac{\partial ^{2}f}{%
\partial S^{2}}\left( \frac{dS}{dt}\right) ^{2}\right\} .  \notag
\end{eqnarray}%
Using the coupled stochastic Langevin equations, the series yields%
\begin{eqnarray}
\frac{df}{dt} &=&\lim_{\epsilon \rightarrow 0}\frac{1}{\epsilon }\left\{
f(t+\epsilon ,S(t+\epsilon ),V(t+\epsilon ))-f(t,S(t),V(t))\right\}
\label{Stochastic GM equation} \\
\frac{df}{dt} &=&\frac{\partial f}{\partial t}+\frac{\partial f}{\partial S}%
\left( \phi S+\sigma SR\right) +\frac{\partial f}{\partial V}\left( \mu
V+\xi VQ\right)  \notag \\
&&+\lim_{\epsilon \rightarrow 0}\frac{\epsilon }{2!}\left\{ 
\begin{array}{c}
\frac{\partial ^{2}f}{\partial S^{2}}\left( \phi S+\sigma SR\right) ^{2}+%
\frac{\partial ^{2}f}{\partial V^{2}}\left( \mu V+\xi VQ\right) ^{2} \\ 
+2\frac{\partial ^{2}f}{\partial S\partial V}\left( \phi S+\sigma SR\right)
\left( \mu V+\xi VQ\right)%
\end{array}%
\right\} .  \notag
\end{eqnarray}%
We expand equation $\left( \ref{Stochastic GM equation}\right) $ and use
equations $\left( \ref{RR noise}\right) $ and $\left( \ref{RQ noise}\right) $%
. Taking the limit $\epsilon \rightarrow 0,$ and separating the Taylor
expansion into the deterministic and stochastic parts, we have%
\begin{eqnarray}
\frac{df}{dt} &=&\frac{\partial f}{\partial t}+\phi S\frac{\partial f}{%
\partial S}+\mu V\frac{\partial f}{\partial V}+\frac{1}{2!}\left\{ 
\begin{array}{c}
\sigma ^{2}S^{2}\frac{\partial ^{2}f}{\partial S^{2}}+\xi ^{2}V^{2}\frac{%
\partial ^{2}f}{\partial V^{2}} \\ 
+2\sigma S\xi V\rho \frac{\partial ^{2}f}{\partial S\partial V}%
\end{array}%
\right\} \\
&&+\sigma S\frac{\partial f}{\partial S}R+\xi V\frac{\partial f}{\partial V}%
Q.  \notag
\end{eqnarray}%
Recall that $V=\sigma ^{2},$we have%
\begin{eqnarray}
\frac{df}{dt} &=&\frac{\partial f}{\partial t}+\phi S\frac{\partial f}{%
\partial S}+\mu V\frac{\partial f}{\partial V}+\frac{1}{2!}\left\{ 
\begin{array}{c}
\sigma ^{2}S^{2}\frac{\partial ^{2}f}{\partial S^{2}}+\xi ^{2}V^{2}\frac{%
\partial ^{2}f}{\partial V^{2}} \\ 
+2\sigma ^{3}S\xi \rho \frac{\partial ^{2}f}{\partial S\partial V}%
\end{array}%
\right\}  \label{time drivative} \\
&&+\sigma S\frac{\partial f}{\partial S}R+\xi V\frac{\partial f}{\partial V}%
Q.  \notag
\end{eqnarray}%
We simplify the equation by the following substitutions%
\begin{eqnarray}
\omega &=&\frac{\partial }{\partial t}+\phi S\frac{\partial }{\partial S}%
+\mu V\frac{\partial }{\partial V}+\frac{1}{2!}\left\{ 
\begin{array}{c}
\sigma ^{2}S^{2}\frac{\partial ^{2}}{\partial S^{2}}+\xi ^{2}V^{2}\frac{%
\partial ^{2}}{\partial V^{2}} \\ 
+2\sigma S\xi V\rho \frac{\partial ^{2}}{\partial S\partial V}%
\end{array}%
\right\} \\
\alpha _{1} &=&\sigma S\frac{\partial f}{\partial S} \\
\alpha _{2} &=&\xi V\frac{\partial }{\partial V}.
\end{eqnarray}%
The equation reduces to%
\begin{equation}
\frac{df}{dt}=\left( \omega +\alpha _{1}R+\alpha _{2}Q\right) f,
\end{equation}%
where $\omega ,$ $\alpha _{1},$ $\alpha _{2}$ are linear operators. We
consider the following self-replicating portfolio%
\begin{equation}
\pi (t)=\theta _{1}f(t)+\theta _{2}S(t),
\end{equation}%
where $\theta _{1}$ and $\theta _{2}$ are the amounts for the associated
option $f$ and security $S.$ The total time derivative of the portfolio
yields%
\begin{eqnarray}
\frac{d\pi }{dt} &=&\theta _{1}\frac{df}{dt}+\theta _{2}\frac{dS}{dt}
\label{portfolio} \\
\frac{d\pi }{dt} &=&\theta _{1}f\omega +\phi \theta _{2}S+\left( \alpha
_{1}R+\alpha _{1}Q\right) \theta _{1}f+\theta _{2}\sigma SR.  \notag
\end{eqnarray}%
By inspection, the total time derivative contains both deterministic and
random terms. To remove the stochastic terms in equation $\left( \ref%
{portfolio}\right) ,$ we use the following matrix relation%
\begin{equation}
\left[ 
\begin{array}{cc}
\alpha _{1} & \sigma \\ 
\alpha _{2} & 0%
\end{array}%
\right] \left[ 
\begin{array}{c}
\theta _{1}f \\ 
\theta _{2}S%
\end{array}%
\right] =0.  \label{Hedging matrix}
\end{equation}%
Equation $\left( \ref{portfolio}\right) $ reduces 
\begin{equation}
\frac{d\pi }{dt}=\theta _{1}f\omega +\phi \theta _{2}S,
\end{equation}%
which is pure deterministic since the stochastic terms containing $R(t)$ and 
$Q(t)$ are removed. The elimination method of the Gaussian noise terms in
equation $\left( \ref{portfolio}\right) $ by using the matrix relation $%
\left( \ref{Hedging matrix}\right) $ is analogous to the hedging technique
used in the case of constant volatility $\sigma \neq \sigma (t)$, i.e. where
the fluctuations of one security is cancelled by another security. The
absence of arbitrage forces the time-derivative of the portfolio to be
directly proportional to the risk-neutral rate $r$, hence 
\begin{equation}
\frac{d\pi }{dt}=r\pi .  \label{risk-neutral}
\end{equation}%
Equating equations $\left( \ref{portfolio}\right) $ and $\left( \ref%
{risk-neutral}\right) $ above yields%
\begin{equation}
\left( \omega -r\right) \theta _{1}f+\left( \phi -r\right) \theta _{2}S=0.
\label{growth equ}
\end{equation}%
The equation $\left( \ref{growth equ}\right) $ can be satisfied by requiring
the following constraint relation%
\begin{equation}
\begin{array}{c}
\left[ 
\begin{array}{cc}
\left( \omega -r\right) & \left( \phi -r\right)%
\end{array}%
\right] =\left[ 
\begin{array}{cc}
\lambda _{1}(t) & \lambda _{2}(t)%
\end{array}%
\right] \left[ 
\begin{array}{cc}
\alpha _{1} & \sigma \\ 
\alpha _{2} & 0%
\end{array}%
\right] ,%
\end{array}
\label{constrained 3}
\end{equation}%
where $\lambda _{1}(t)$ and $\lambda _{2}(t)$ are arbitrary. Writing
equation $\left( \ref{constrained 3}\right) $ in component forms%
\begin{eqnarray}
\omega -r &=&\lambda _{1}(t)\alpha _{1}+\lambda _{2}(t)\alpha _{2} \\
\phi -r &=&\lambda _{1}(t)\sigma .
\end{eqnarray}%
Substituting for $\lambda _{1}(t),$ equation $\left( \ref{growth equ}\right) 
$ becomes%
\begin{equation}
\left( \omega -r\right) f=\left\{ \left( \frac{\phi -r}{\sigma }\right)
\alpha _{1}+\lambda _{2}(t)\alpha _{2}\right\} f.
\end{equation}%
Upon re-substitution, we obtain the Merton-Garman equation%
\begin{eqnarray}
&&\frac{\partial f}{\partial t}+\frac{1}{2!}\left\{ \sigma ^{2}S^{2}\frac{%
\partial ^{2}f}{\partial S^{2}}+\xi ^{2}V^{2}\frac{\partial ^{2}f}{\partial
V^{2}}+2\sigma ^{3}S\xi \rho \frac{\partial ^{2}f}{\partial S\partial V}%
\right\} -rf \\
&=&-rS\frac{\partial f}{\partial S}+\left( \lambda _{2}(t)\xi -\mu \right) V%
\frac{\partial f}{\partial S}.  \notag
\end{eqnarray}%
For a large class of problems, Hull and White \cite{Hull} argued that we can
redefine 
\begin{equation}
\lambda _{2}(t)\xi -\mu =-\overline{\mu }.  \label{Hull}
\end{equation}%
The Merton-Garman equation becomes%
\begin{eqnarray}
&&\frac{\partial f}{\partial t}+\frac{1}{2!}\left\{ \sigma ^{2}S^{2}\frac{%
\partial ^{2}f}{\partial S^{2}}+\xi ^{2}V^{2}\frac{\partial ^{2}f}{\partial
V^{2}}+2\sigma ^{3}S\xi \rho \frac{\partial ^{2}f}{\partial S\partial V}%
\right\} -rf  \label{Stochastic GM equation 3} \\
&=&-rS\frac{\partial f}{\partial S}-\overline{\mu }\sigma ^{2}\frac{\partial
f}{\partial S}.  \notag
\end{eqnarray}%
This equation is valid for any security derivative $f$ with stochastic
volatility.

\section{Extra-Dimensional Theories and Application}

This section illustrates one of the many tools of theoretical physics in
analyzing problems associated with option trading. The idea of extra
dimensions dated back to Kaluza \ \cite{Kaluza} at the early 1920's. In this
era, Maxwell's beautiful unified electromagnetic theory had inspired
Einstein to unify space and time into spacetime. The spacetime unification
and other postulates had allowed Einstein to formulate the general theory of
relativity. Follow the same guiding principle, Kaluza was successfully able
to unify gravity and electromagnetism by postulating an extra spacetime
dimension. The electromagnetic theory was further enhanced by Klein \cite%
{Klein} in the mid 1920's. The enhanced unified theory of gravity and
electromagnetism is known today as Kaluza-Klein theory.

To illustrate the method, we consider a five dimensional $5D$ spacetime with
the following flat metric: $\eta _{AB}=(\eta _{\mu \nu },\zeta ),$ where $%
\eta _{\mu \nu }=(-1,\delta _{ik})$ is four dimensional (4D) flat Minkowski
metric. The Capital Latin indices run in 5D spacetime as $A,B,..=0,1,2,3,5,$
Greek indices run in 4D space-time $\mu ,\nu ,..=0,1,2,3,$and small Latin
indices in 3D Euclidean space $i,j,..=1,2,3.$ The symbol $\epsilon =\pm 1$
(-1 for a time-like extra dimension, +1 for a space-like extra dimension)
represents the signature of the extra-dimensions. To obtain the proper 5D
Klein-Gordon equation, we have to reanalyze the 5D energy-momentum relation.
Let us for a moment define the 5-position vector of the particle as $%
x^{A}=\left( x^{0},x^{1},x^{2},x^{3},x^{5}\right) =\left(
ct,x^{1},x^{2},x^{3},x^{5}\right) $, with the extra dimension being
uncompactified and time-dependent $x^{5}=x^{5}(t)$. The 5-velocity is
defined by 
\begin{equation}
U^{A}=\frac{dx^{A}}{d\tau }=\gamma \frac{dx^{A}}{dt}=\gamma \left( c,%
\overrightarrow{v},\frac{dx^{5}}{dt}\right) ,
\end{equation}%
where $\gamma $ is the usual Lorentz factor from the relation from special
relativity, $t=\gamma \tau ,$ and $\tau $ is the proper time$.$ The
5D-momentum is defined by 
\begin{equation}
P^{A}=mU^{A}=\gamma m\left( c,\overrightarrow{v},\frac{dx^{5}}{dt}\right) .
\end{equation}%
Analogous to 4D, the 5D energy-momentum relation can be obtained by 
\begin{eqnarray}
P^{A}P_{A} &=&\gamma ^{2}m^{2}\left( -c^{2}+\left\vert \overrightarrow{v}%
\right\vert ^{2}+\epsilon \left( \frac{dx^{5}}{dt}\right) ^{2}\right) \\
&=&\frac{-m^{2}c^{2}}{1-\frac{v^{2}}{c^{2}}}+\frac{m^{2}v^{2}}{1-\frac{v^{2}%
}{c^{2}}}+\epsilon \gamma ^{2}m^{2}\left( \frac{dx^{5}}{dt}\right) ^{2} 
\notag \\
&=&-m^{2}c^{2}+\epsilon \gamma ^{2}m^{2}\left( \frac{dx^{5}}{dt}\right) ^{2}.
\notag
\end{eqnarray}%
If $x^{5}$ is time-dependent, led to a non-invariant expression. Therefore $%
x^{5}$ must be time-independent, so that we can have a vanishing term 
\begin{equation}
\epsilon \gamma ^{2}m^{2}\left( \frac{dx^{5}}{dt}\right) ^{2}=0.
\end{equation}%
To obtain a proper 5D Klein-Gordon equation, we must rewrite the above
expression by absorbing the Lorentz factor with the extra component%
\begin{equation}
\epsilon m^{2}\left( \frac{d(\gamma x^{5})}{dt}\right) ^{2}=\epsilon
m^{2}\left( \frac{d(\overline{x}^{5})}{dt}\right) ^{2}=0,
\end{equation}%
where $\overline{x}^{5}=\gamma x^{5},$ the relativistic extra component of
the 5D position vector $\overline{x}^{5}$. The reason for redefining the
extra dimension this way will apparent when we examine the energy from the
higher dimensional energy-momentum relation. The 5D Klein-Gordon equation:%
\begin{equation}
(\square _{5}-m^{2}c^{2})\Phi =0,
\end{equation}%
where $\square _{5}=\eta ^{AB}\partial _{B}\partial _{A}=\eta ^{\mu \nu
}\partial _{\nu }\partial _{\mu }+\epsilon \partial _{5}^{2}.$ We then
compactified the extra spacelike dimension and let it be spanned by is to be
spanned by the relativistic component, $\overline{x}^{5}.$ The momentum
along the extra dimension, $\overline{x}^{5},$ is quantized by $\frac{n}{L}$%
, where $L$ is the radius of the compactified extra dimension, and $%
n\epsilon Z.$ This choice of quantization led the following periodic
condition along the relativistic component $\overline{x}^{5}$%
\begin{equation}
\Phi (x^{\mu },\overline{x}^{5})=\Phi (x^{\mu },\overline{x}^{5}+2\pi L).
\end{equation}%
The proper 5D Klein-Gordon equation can be written a manifestly in 4D by
considering the following field decomposition:%
\begin{equation}
\Phi (x^{\mu },\overline{x}^{5})=\dsum\limits_{n}\Phi _{n}(x^{\mu })\exp
\left( \frac{in\overline{x}^{5}}{L}\right) ,n=0,\pm 1,\pm 2,....
\end{equation}%
Using the above field decomposition and the 5D Klein-Gordon equation becomes
\ 
\begin{equation}
(\square -\epsilon \frac{\gamma ^{2}n^{2}}{\left( 2\pi L\right) ^{2}}%
-m^{2}c^{2})\Phi _{n}(x^{\mu })=0.
\end{equation}%
For space-like extra dimension, $\epsilon =1,$ and letting the effective
mass be $m_{eff}=\frac{\gamma ^{2}n^{2}}{\left( 2\pi L\right) ^{2}}%
+m^{2}c^{2}.$ The $5D$ Klein-Gordon equation reduces to the $4D$
Klein-Gordon equation%
\begin{equation}
(\square -m_{eff})\Phi _{n}(x^{\mu })=0.  \label{Modified KG equation}
\end{equation}%
The effect of the compactified extra dimensions and the associated periodic
boundary conditions is equivalent to an increase in the mass of the
propagating particles, i.e. the modified Klein-Gordon equation $\left( \ref%
{Modified KG equation}\right) $.

\section{Merton-Garman Equation via Extra Dimensional Approach}

In this section, with the appropriate constraints, the Merton-Garman
equation can be derived directly from a higher dimensional space or
superspace. Furthermore, the stochastic or random nature of the security $%
S(t)$ and volatility $V(t)$ are seen as the effects of the extra dimensions.
Mathematically, $S(t)$ and $V(t)$ are being projected from the higher
manifold or supermanifold by a higher dimensional operators. The time
evolution $S(t)$ and $V(t)$ are governed by super stochastic differential
equation.

Consider an option of a security to assume the following form 
\begin{equation}
f=f_{5}=f(t,S(t),V(t),x^{5}(t),x^{6}(t)),
\end{equation}%
where $t,$ $S(t),$ and $V(t)$ are the usual time, security and volatility
respectively and the coordinate functions, $x^{5}(t)$ and $x^{6}(t),$ are
referred to simply as the coordinates of the extra dimensions. The function $%
f_{5}$ is a real-valued function which maps an element of $R^{5}\subset E^{5}
$ into the real $R,$ where $E^{5}$ is $5D$ Euclidean space%
\begin{equation}
f_{5}:R^{5}\rightarrow R.
\end{equation}%
Consider a supercoordinate or $5D$ vector to represent a point in $R^{5}$ 
\begin{eqnarray}
\mathbf{z}^{A} &=&(z^{2},z^{3},z^{4},z^{5},z^{6})=(t,S,V,x^{5},x^{6}) \\
\mathbf{z}^{A} &=&t\widehat{t}+S\widehat{s}+V\widehat{v}+x^{5}\widehat{x}%
_{5}+x^{6}\widehat{x}_{6},
\end{eqnarray}%
where $A=2,3,4,5,6$. The unit vectors 
\begin{equation}
\widehat{t}=\left( 
\begin{array}{c}
1 \\ 
0 \\ 
0 \\ 
0 \\ 
0%
\end{array}%
\right) ,\widehat{s}=\left( 
\begin{array}{c}
0 \\ 
1 \\ 
0 \\ 
0 \\ 
0%
\end{array}%
\right) ,\widehat{v}=\left( 
\begin{array}{c}
0 \\ 
0 \\ 
1 \\ 
0 \\ 
0%
\end{array}%
\right) ,\widehat{x}_{5}=\left( 
\begin{array}{c}
0 \\ 
0 \\ 
0 \\ 
1 \\ 
0%
\end{array}%
\right) ,\widehat{x}_{6}=\left( 
\begin{array}{c}
0 \\ 
0 \\ 
0 \\ 
0 \\ 
1%
\end{array}%
\right) .
\end{equation}%
We could also write the supercoordinate $\mathbf{z}^{A}$ in matrix notation
as 
\begin{eqnarray}
\mathbf{z}^{A} &=&t\widehat{t}+S\widehat{s}+V\widehat{v}+x^{5}\widehat{x}%
_{5}+x^{6}\widehat{x}_{6} \\
\mathbf{z}^{A} &=&t\left( 
\begin{array}{c}
1 \\ 
0 \\ 
0 \\ 
0 \\ 
0%
\end{array}%
\right) +S\left( 
\begin{array}{c}
0 \\ 
1 \\ 
0 \\ 
0 \\ 
0%
\end{array}%
\right) +V\left( 
\begin{array}{c}
0 \\ 
0 \\ 
1 \\ 
0 \\ 
0%
\end{array}%
\right) +x^{5}\left( 
\begin{array}{c}
0 \\ 
0 \\ 
0 \\ 
1 \\ 
0%
\end{array}%
\right) +x^{6}\left( 
\begin{array}{c}
0 \\ 
0 \\ 
0 \\ 
0 \\ 
1%
\end{array}%
\right)  \\
\mathbf{z}^{A} &=&\left( 
\begin{array}{ccccc}
1 & 0 & 0 & 0 & 0 \\ 
0 & 1 & 0 & 0 & 0 \\ 
0 & 0 & 1 & 0 & 0 \\ 
0 & 0 & 0 & 1 & 0 \\ 
0 & 0 & 0 & 0 & 1%
\end{array}%
\right) \left( 
\begin{array}{c}
t \\ 
S \\ 
V \\ 
x^{5} \\ 
x^{6}%
\end{array}%
\right) =I_{5}z^{A}.
\end{eqnarray}%
The magnitude of the $5D$ vector is%
\begin{equation}
\left\vert z^{A}\right\vert =\left\vert I_{5}z^{A}\right\vert =\sqrt{%
t^{2}+S^{2}+V^{2}+\left( x^{5}\right) ^{2}+\left( x^{6}\right) ^{2}}.
\end{equation}%
We can write $z^{A}$ in term of arbitrary unit vector by first finding the
unit vector in the $z$ direction to be%
\begin{equation}
\widehat{z}=\frac{\mathbf{z}^{A}}{\left\vert z^{A}\right\vert }=\frac{1}{%
\sqrt{t^{2}+S^{2}+V^{2}+\left( x^{5}\right) ^{2}+\left( x^{6}\right) ^{2}}}%
\left( 
\begin{array}{c}
t \\ 
S \\ 
V \\ 
x^{5} \\ 
x^{6}%
\end{array}%
\right) .
\end{equation}%
Taking the total differential of $\mathbf{z}^{A}$ to first order, we have%
\begin{eqnarray}
d\left( \mathbf{z}^{A}\right)  &=&\dsum\limits_{B=2}^{6}\frac{\partial 
\mathbf{z}^{A}}{\partial z^{B}}dz^{B}  \notag \\
d\left( \mathbf{z}^{A}\right)  &=&\frac{\partial \mathbf{z}^{A}}{\partial t}%
dt+\frac{\partial \mathbf{z}^{A}}{\partial S}dS+\frac{\partial \mathbf{z}^{A}%
}{\partial V}dV+\frac{\partial \mathbf{z}^{A}}{\partial x^{5}}dx^{5}+\frac{%
\partial \mathbf{z}^{A}}{\partial x^{6}}  \notag \\
d\left( \mathbf{z}^{A}\right)  &=&dt\widehat{t}+dS\widehat{s}+dV\widehat{v}%
+dx^{5}\widehat{x}_{5}+dx^{6}\widehat{x}_{6}.
\end{eqnarray}%
Recall that $f_{5}:R^{5}\rightarrow R,$ and define by $y=f_{5}(z^{A}).$ By
inspection, taking a total differential of $f_{5}(\mathbf{z}^{A})$ is
equivalent to a mapping from 
\begin{equation}
df_{5}(\mathbf{z}^{A}):R\rightarrow R.
\end{equation}%
The mapping gives%
\begin{eqnarray}
df_{5}\left( \mathbf{z}^{A}\right)  &=&\dsum\limits_{B=2}^{6}\frac{\partial
f_{5}\left( \mathbf{z}^{A}\right) }{\partial z^{B}}dz^{B}  \notag \\
df_{5}\left( \mathbf{z}^{A}\right)  &=&\frac{\partial f_{5}}{\partial t}dt+%
\frac{\partial f_{5}}{\partial S}dS+\frac{\partial f_{5}}{\partial V}dV+%
\frac{\partial f_{5}}{\partial x^{5}}dx^{5}+\frac{\partial f_{5}}{\partial
x^{6}}dx^{6}  \notag \\
df_{5}\left( \mathbf{z}^{A}\right)  &=&\frac{\partial f_{5}}{\partial t}%
\widehat{t}\cdot \widehat{t}dt+\frac{\partial f_{5}}{\partial S}\widehat{s}%
\cdot \widehat{s}dS+\frac{\partial f_{5}}{\partial V}\widehat{v}\cdot 
\widehat{v}dV  \notag \\
&&+\frac{\partial f_{5}}{\partial x^{5}}\widehat{x}_{5}\cdot \widehat{x}%
_{5}dx^{5}+\frac{\partial f_{5}}{\partial x^{6}}\widehat{x}_{6}\cdot 
\widehat{x}_{6}dx^{6}  \notag \\
df_{5}\left( \mathbf{z}^{A}\right)  &=&grad_{5}\left( f_{5}\right) \cdot d%
\mathbf{z}^{A},
\end{eqnarray}%
where 
\begin{equation*}
grad_{5}\left( {}\right) =\frac{\partial }{\partial t}\widehat{t}+\frac{%
\partial }{\partial S}\widehat{s}+\frac{\partial }{\partial V}\widehat{v}+%
\frac{\partial }{\partial x^{5}}\widehat{x}_{5}+\frac{\partial }{\partial
x^{6}}\widehat{x}_{6}.
\end{equation*}%
In matrix notation, we have 
\begin{eqnarray}
df_{5}\left( \mathbf{z}^{A}\right)  &=&\left( 
\begin{array}{ccccc}
\frac{\partial f_{5}}{\partial t} & \frac{\partial f_{5}}{\partial s} & 
\frac{\partial f_{5}}{\partial V} & \frac{\partial f_{5}}{\partial x^{5}} & 
\frac{\partial f_{5}}{\partial x^{6}}%
\end{array}%
\right) \cdot \left( 
\begin{array}{c}
dt \\ 
dS \\ 
dV \\ 
dx^{5} \\ 
dx^{6}%
\end{array}%
\right)  \\
df_{5}\left( \mathbf{z}^{A}\right)  &=&grad_{5}\left( f_{5}\right) \cdot d%
\mathbf{z}^{A}.  \notag
\end{eqnarray}%
In order to project pertinent information from superspace onto ordinary
space, we define a mapping or projection by $\Pi :R^{5}\rightarrow R^{3},$
such that 
\begin{eqnarray}
\Pi df_{5}\left( \mathbf{z}^{A}\right)  &=&\Pi \left[ \frac{\partial f_{5}}{%
\partial t}dt+\frac{\partial f_{5}}{\partial S}dS+\frac{\partial f_{5}}{%
\partial V}dV+\frac{\partial f_{5}}{\partial x^{5}}dx^{5}+\frac{\partial
f_{5}}{\partial x^{6}}dx^{6}\right]   \notag \\
df_{5}\left( \Pi \mathbf{z}^{A}\right)  &=&\frac{\partial f_{5}}{\partial t}%
dt+\frac{\partial f_{5}}{\partial S}\left( dS+\frac{\partial S}{\partial
x^{5}}dx^{5}\right) +\frac{\partial f_{5}}{\partial V}\left( dV+\frac{%
\partial V}{\partial x^{6}}dx^{6}\right)   \label{Mapping} \\
df_{5}\left( \mathbf{z}^{\mu }\right)  &=&\frac{\partial f_{5}}{\partial t}%
dt+\frac{\partial f_{5}}{\partial S}\left( dS+\frac{\partial S}{\partial
x^{5}}dx^{5}\right) +\frac{\partial f_{5}}{\partial V}\left( dV+\frac{%
\partial V}{\partial x^{6}}dx^{6}\right) ,  \notag
\end{eqnarray}%
and requiring the following constrained equations 
\begin{eqnarray}
\frac{\partial f_{5}}{\partial S}\left( dS+\frac{\partial S}{\partial x^{5}}%
dx^{5}\right)  &=&\left( -i\frac{\partial f_{5}}{\partial x^{5}}\right) 
\frac{\partial }{\partial S}, \\
\frac{\partial f_{5}}{\partial V}dV+\frac{\partial f}{\partial x^{6}}dx^{6}
&=&\left( -i\frac{\partial f_{5}}{\partial x^{6}}\right) \frac{\partial }{%
\partial V}.
\end{eqnarray}%
Writing equation $\left( \ref{Mapping}\right) $ in matrix form allows us to
see that important information has been projected from superspace by the
mapping $\Pi ,$%
\begin{eqnarray}
df_{5}\left( \mathbf{z}^{\mu }\right)  &=&\left( 
\begin{array}{ccc}
\frac{\partial f_{5}}{\partial t} & \frac{\partial f_{5}}{\partial s} & 
\frac{\partial f_{5}}{\partial V}%
\end{array}%
\right) \cdot \left( 
\begin{array}{c}
dt \\ 
dS+\frac{\partial S}{\partial x^{5}}dx^{5} \\ 
dV+\frac{\partial V}{\partial x^{6}}dx^{6}%
\end{array}%
\right)  \\
df_{5}\left( \mathbf{z}^{\mu }\right)  &=&\left( 
\begin{array}{ccc}
\frac{\partial f_{5}}{\partial t} & \frac{\partial }{\partial s} & \frac{%
\partial }{\partial V}%
\end{array}%
\right) \cdot \left( 
\begin{array}{c}
dt \\ 
\left( -i\frac{\partial }{\partial x^{5}}\right) f_{5} \\ 
\left( -i\frac{\partial }{\partial x^{6}}\right) f_{5}%
\end{array}%
\right) ,  \notag
\end{eqnarray}%
where $\mu =2,3,4$ is the index of the ordinary coordinates. We require that
the coordinate functions of the extra dimensions to be
ordinary-coordinate-independent, $x^{5}\neq x^{5}(S,V)$ and $x^{6}\neq
x^{6}(S,V).$

Algebraically, we can demonstrate the projections from superspace onto
ordinary space by writing down the second-order Taylor series expansion for
the total differential of the option $\Pi df_{5}\left( \mathbf{z}^{A}\right) 
$%
\begin{eqnarray}
\Pi df_{5}\left( \mathbf{z}^{A}\right)  &=&df_{5}\left( \Pi \mathbf{z}%
^{A}\right) =\lim_{\epsilon \rightarrow 0}\frac{\Pi }{\epsilon }\left\{ 
\begin{array}{c}
f_{5}(t+\epsilon ,S(t+\epsilon ),V(t+\epsilon ),x^{5}(t+\epsilon
),x^{6}(t+\epsilon )) \\ 
-f_{5}(t,S(t),V(t),x^{5}(t),x^{6}(t))%
\end{array}%
\right\} dt  \notag \\
df_{5}\left( \mathbf{z}^{\mu }\right)  &=&\Pi \{\frac{\partial f_{5}}{%
\partial t}dt+\frac{\partial f_{5}}{\partial S}dS+\frac{\partial f_{5}}{%
\partial V}dV+\frac{\partial f_{5}}{\partial x^{5}}dx^{5}+\frac{\partial
f_{5}}{\partial x^{6}}dx^{6} \\
+\lim_{\epsilon \rightarrow 0}\frac{\epsilon }{2!} &&\left\{ 
\begin{array}{c}
\frac{\partial }{\partial S}\left( \frac{\partial f_{5}}{\partial S}%
dS\right) dS+\frac{\partial }{\partial V}\left( \frac{\partial f_{5}}{%
\partial S}dS\right) dV+\frac{\partial }{\partial x^{5}}\left( \frac{%
\partial f_{5}}{\partial S}dS\right) dx^{5} \\ 
+\frac{\partial }{\partial x^{6}}\left( \frac{\partial f_{5}}{\partial S}%
dS\right) dx^{6}+\frac{\partial }{\partial S}\left( \frac{\partial f_{5}}{%
\partial V}dV\right) dS+\frac{\partial }{\partial V}\left( \frac{\partial
f_{5}}{\partial V}dV\right) dV \\ 
+\frac{\partial }{\partial x^{5}}\left( \frac{\partial f_{5}}{\partial V}%
dV\right) dx^{5}+\frac{\partial }{\partial x^{6}}\left( \frac{\partial f_{5}%
}{\partial V}dV\right) dx^{6}+\frac{\partial }{\partial S}\left( \frac{%
\partial f_{5}}{\partial x^{5}}dx^{5}\right) dS \\ 
+\frac{\partial }{\partial V}\left( \frac{\partial f_{5}}{\partial x^{5}}%
dx^{5}\right) dV+\frac{\partial }{\partial x^{5}}\left( \frac{\partial f_{5}%
}{\partial x^{5}}dx^{5}\right) dx^{5}+\frac{\partial }{\partial x^{6}}\left( 
\frac{\partial f_{5}}{\partial x^{5}}dx^{5}\right) dx^{6} \\ 
\frac{\partial }{\partial S}\left( \frac{\partial f_{5}}{\partial x^{6}}%
dx^{6}\right) dS+\frac{\partial }{\partial V}\left( \frac{\partial f_{5}}{%
\partial x^{6}}dx^{6}\right) dV+\frac{\partial }{\partial x^{5}}\left( \frac{%
\partial f_{5}}{\partial x^{6}}dx^{6}\right) dx^{5} \\ 
+\frac{\partial }{\partial x^{6}}\left( \frac{\partial f_{5}}{\partial x^{6}}%
dx^{6}\right) dx^{6}%
\end{array}%
\right\} \}.  \notag
\end{eqnarray}%
Recall from section 3, we can write a field decomposition of the option
value $f_{5}$ as%
\begin{equation}
f_{5}(t,S(t),V(t),x^{5},x^{6})=\dsum\limits_{n,m=0}^{\infty }\widehat{f}%
_{3}(t,S(t),V(t))_{nm}\exp i\left( 
\begin{array}{c}
\frac{n}{R_{5}}\cdot x^{5} \\ 
+\frac{m}{R_{5}}\cdot x^{6}%
\end{array}%
\right) ,
\end{equation}%
and is subjected to the periodic condition%
\begin{equation*}
f_{5}(t,S(t),V(t),x^{5},x^{6})=f_{5}(t,S(t),V(t),x^{5}(t)+2n\pi
R_{5},x^{6}(t)+2m\pi R_{6}),
\end{equation*}%
where $\widehat{f}_{3}(t,S(t),V(t))_{nm}$ is the Fourier expansion
coefficients, and $(n,m)\epsilon Z$ are integers with radii , $R_{5}$ and $%
R_{6},$ are the compactified extra dimensions in the, $x^{5}$ and $x^{6}$
directions, respectively. The associated momenta in the extra dimensional
directions are quantized as follows 
\begin{eqnarray}
&&p_{5=}\frac{n}{R_{5}}, \\
&&p_{5=}\frac{n}{R_{5}}.
\end{eqnarray}%
In order to simplify the truncated Taylor series, we assume the following
constrained relations by demanding that%
\begin{equation}
\frac{\partial f_{5}}{\partial S}dS+\frac{\partial f_{5}}{\partial x^{5}}%
dx^{5}=\left( -i\frac{\partial f_{5}}{\partial x^{5}}\right) \frac{\partial 
}{\partial S}  \label{constrained 1}
\end{equation}%
and 
\begin{equation}
\frac{\partial f_{5}}{\partial V}dV+\frac{\partial f_{5}}{\partial x^{6}}%
dx^{6}=\left( -i\frac{\partial f_{5}}{\partial x^{6}}\right) \frac{\partial 
}{\partial V},  \label{constrained 2}
\end{equation}%
where we define the superprojection operators $P_{5}=$ $-i\frac{\partial }{%
\partial x^{5}}$ and $P_{6}=$ $-i\frac{\partial }{\partial x^{6}}.$ Applying
the projection operators on the option $f_{5}$, we have the following
equations 
\begin{eqnarray}
P_{5}f_{5} &=&\left( -i\frac{\partial }{\partial x^{5}}\right) f_{5}
\label{evolu-S(t)} \\
&=&\left( -i\frac{\partial }{\partial x^{5}}\right) \left\{
\dsum\limits_{n,m=0}^{\infty }\widehat{f}_{3}(t,S(t),V(t))_{nm}\exp i\left( 
\begin{array}{c}
\frac{n}{R_{5}}\cdot x^{5} \\ 
+\frac{m}{R_{6}}\cdot x^{6}%
\end{array}%
\right) \right\}   \notag \\
&=&\frac{n}{R_{5}}f_{5},  \notag
\end{eqnarray}%
similarly%
\begin{eqnarray}
P_{6}f_{5} &=&\left( -i\frac{\partial }{\partial x^{6}}\right) f_{5}
\label{evolu-V(t)} \\
&=&-i\frac{\partial }{\partial x^{6}}\left\{ \dsum\limits_{n,m=0}^{\infty }%
\widehat{f_{3}}(t,S(t),V(t))_{nm}\exp i\left( 
\begin{array}{c}
\frac{n}{R_{5}}\cdot x^{5} \\ 
+\frac{m}{R_{6}}\cdot x^{6}%
\end{array}%
\right) \right\}   \notag \\
&=&\frac{m}{R_{6}}f_{5}.  \notag
\end{eqnarray}%
The equations $\left( \ref{evolu-S(t)}\right) $ and $\left( \ref{evolu-V(t)}%
\right) $ give us the explicit quantized forms of the higher dimensional
operators as%
\begin{equation}
P_{5}=\frac{n}{R_{5}},  \label{higher dimen oper 1}
\end{equation}%
and 
\begin{equation}
P_{6}=\frac{m}{R_{6}},  \label{higher dimen oper 2}
\end{equation}%
where $m,n\epsilon Z$ are integers. More importantly, equations $\left( \ref%
{higher dimen oper 1}\right) $ and $\left( \ref{higher dimen oper 2}\right) $
control the stochastic contributions coming from the higher dimensional
subspace of superspace. The constrained equations $\left( \ref{constrained 1}%
\right) $ and $\left( \ref{constrained 2}\right) $ become 
\begin{eqnarray}
\frac{\partial f_{5}}{\partial S}dS+\frac{\partial f_{5}}{\partial x^{5}}%
dx^{5} &=&\left( -i\frac{\partial f_{5}}{\partial x^{5}}\right) \frac{%
\partial }{\partial S}=\frac{n}{R_{5}}\frac{\partial f_{5}}{\partial S} \\
\frac{\partial f_{5}}{\partial V}dV+\frac{\partial f_{5}}{\partial x^{6}}%
dx^{6} &=&\left( -i\frac{\partial f_{5}}{\partial x^{6}}\right) \frac{%
\partial }{\partial V}=\frac{m}{R_{6}}\frac{\partial f_{5}}{\partial V}.
\end{eqnarray}%
Recall that our assumed supermanifold such that coordinate functions of the
extra dimensions to be ordinary-coordinate-independent, $x^{5}\neq x^{5}(S,V)
$ and $x^{6}\neq x^{6}(S,V),$ then we have%
\begin{eqnarray}
\frac{\partial f_{5}}{\partial S}dS+\frac{\partial f_{5}}{\partial x^{5}}%
dx^{5} &=&\frac{n}{R_{5}}\frac{\partial f_{5}}{\partial S}  \notag \\
\frac{\partial f_{5}}{\partial S}\left( dS+\frac{\partial S}{\partial f}%
\cdot \frac{\partial f}{\partial x^{5}}dx^{5}\right)  &=&\frac{n}{R_{5}}%
\frac{\partial f_{5}}{\partial S}  \notag \\
\frac{\partial f_{5}}{\partial S}\left( dS+\frac{\partial S}{\partial x^{5}}%
dx^{5}\right)  &=&\frac{n}{R_{5}}\frac{\partial f_{5}}{\partial S}.
\label{orthogonality}
\end{eqnarray}%
\begin{equation}
\frac{\partial S}{\partial x^{5}}=0.  \label{cons 4}
\end{equation}%
Equation $\left( \ref{orthogonality}\right) $ reduces to 
\begin{equation*}
\frac{\partial f_{5}}{\partial S}dS=\frac{n}{R_{5}}\frac{\partial f_{5}}{%
\partial S}.
\end{equation*}%
This tells us the radius of \ the compactified radius must be $R_{5}=dS^{-1}=%
\left[ \left( \phi S+\sigma SR\right) dt\right] ^{-1}.$ Similarly, the
compactified radius in the $\ x^{6}$ direction is $R_{6}=dV^{-1}=\left[
\left( \mu V+\xi VQ\right) dt\right] ^{-1}.$ The truncated Taylor series
simplifies to 
\begin{eqnarray}
df_{5}\left( \mathbf{z}^{\mu }\right)  &=&\lim_{\epsilon \rightarrow 0}\frac{%
\Pi }{\epsilon }\left\{ 
\begin{array}{c}
f_{5}(t+\epsilon ,S(t+\epsilon ),V(t+\epsilon ),x^{5}(t+\epsilon
),x^{6}(t+\epsilon )) \\ 
-f_{5}(t,S(t),V(t),x^{5}(t),x^{6}(t))%
\end{array}%
\right\} dt  \notag \\
df_{5}\left( \mathbf{z}^{\mu }\right)  &=&\frac{\partial f_{5}}{\partial t}%
dt+\left( -i\frac{\partial f_{5}}{\partial x^{5}}\right) \frac{\partial }{%
\partial S}+\left( -i\frac{\partial f_{5}}{\partial x^{6}}\right) \frac{%
\partial }{\partial V}  \notag \\
&&+\lim_{\epsilon \rightarrow 0}\frac{\epsilon }{2!}\left\{ 
\begin{array}{c}
\frac{\partial ^{2}f_{5}}{\partial S^{2}}dS^{2}+2\frac{\partial ^{2}f_{5}}{%
\partial S\partial V}dSdV+\frac{\partial ^{2}f_{5}}{\partial V^{2}}dV^{2} \\ 
+\frac{\partial ^{2}f_{5}}{\left( \partial x^{5}\right) ^{2}}\left(
dx^{5}\right) ^{2}+\frac{\partial ^{2}f_{5}}{\left( \partial x^{6}\right)
^{2}}\left( dx^{6}\right) ^{2} \\ 
+2\frac{\partial ^{2}f_{5}}{\partial x^{5}\partial S}dSdx^{5}+2\frac{%
\partial ^{2}f_{5}}{\partial x^{6}\partial S}dSdx^{6} \\ 
+2\frac{\partial ^{2}f_{5}}{\partial x^{5}\partial V}dVdx^{5}+2\frac{%
\partial ^{2}f_{5}}{\partial x^{6}\partial V}dVdx^{6} \\ 
+2\frac{\partial ^{2}f_{5}}{\partial x^{5}\partial x^{6}}dx^{5}dx^{6}%
\end{array}%
\right\} .
\end{eqnarray}%
Upon further simplification,we have the following relations 
\begin{eqnarray}
\frac{\partial ^{2}f_{5}}{\partial S^{2}}dS^{2}+2\frac{\partial ^{2}f_{5}}{%
\partial x^{6}\partial S}dSdx^{6} &=&\frac{\partial }{\partial S}dS\left( 
\frac{\partial f_{5}}{\partial S}dS+2\frac{\partial f_{5}}{\partial x^{6}}%
dx^{6}\right)  \\
&=&\frac{\partial }{\partial S}dS\frac{\partial f_{5}}{\partial S}\left( dS+%
\frac{\partial S}{\partial x^{6}}dx^{6}\right)   \notag \\
&=&\frac{\partial ^{2}f_{5}}{\partial S^{2}}dS^{2},  \notag
\end{eqnarray}%
\begin{eqnarray}
&&\frac{\partial ^{2}f_{5}}{\partial S^{2}}dS^{2}+2\frac{\partial ^{2}f_{5}}{%
\partial x^{5}\partial S}dSdx^{5} \\
&=&\frac{\partial }{\partial S}dS\left( \frac{\partial f_{5}}{\partial S}dS+2%
\frac{\partial f_{5}}{\partial x^{5}}dx^{5}\right)   \notag \\
&=&\frac{\partial }{\partial S}dS\left( \frac{\partial f_{5}}{\partial S}dS+%
\frac{\partial f_{5}}{\partial x^{5}}dx^{5}\right) +\frac{\partial ^{2}f_{5}%
}{\partial S\partial x^{5}}dSdx^{5}  \notag \\
&=&\frac{\partial }{\partial S}dS\frac{\partial f_{5}}{\partial S}\left( dS+%
\frac{\partial S}{\partial x^{5}}dx^{5}\right) +\frac{\partial ^{2}f_{5}}{%
\partial S\partial x^{5}}dSdx^{5}  \notag \\
&=&\frac{\partial ^{2}}{\partial S^{2}}dS\left( -i\frac{\partial f_{5}}{%
\partial x^{5}}\right) +\frac{\partial ^{2}f_{5}}{\partial S\partial x^{5}}%
dSdx^{5},  \notag
\end{eqnarray}%
similarly,%
\begin{eqnarray}
&&\frac{\partial ^{2}f_{5}}{\partial V^{2}}dV^{2}+2\frac{\partial ^{2}f_{5}}{%
\partial x^{5}\partial V}dVdx^{5}+2\frac{\partial ^{2}f_{5}}{\partial
x^{6}\partial V}dVdx^{6} \\
&=&\frac{\partial ^{2}f_{5}}{\partial V^{2}}dV^{2}+2\frac{\partial ^{2}f_{5}%
}{\partial x^{6}\partial V}dVdx^{6}  \notag \\
&=&\frac{\partial }{\partial S}dS\frac{\partial f_{5}}{\partial S}\left( dS+%
\frac{\partial S}{\partial x^{6}}dx^{6}\right)   \notag \\
&=&\frac{\partial ^{2}}{\partial V^{2}}dV\left( -i\frac{\partial f_{5}}{%
\partial x^{6}}\right) +\frac{\partial ^{2}f_{5}}{\partial V\partial x^{6}}%
dVdx^{6}.  \notag
\end{eqnarray}%
In terms of the superprojection operators, the truncated Taylor series
becomes%
\begin{eqnarray}
df_{5}\left( \mathbf{z}^{\mu }\right)  &=&\lim_{\epsilon \rightarrow 0}\frac{%
\Pi }{\epsilon }\left\{ 
\begin{array}{c}
f_{5}(t+\epsilon ,S(t+\epsilon ),V(t+\epsilon ),x^{5}(t+\epsilon
),x^{6}(t+\epsilon )) \\ 
-f_{5}(t,S(t),V(t),x^{5}(t),x^{6}(t))%
\end{array}%
\right\} dt \\
df_{5}\left( \mathbf{z}^{\mu }\right)  &=&\Pi \{\frac{\partial f_{5}}{%
\partial t}dt+\frac{\partial }{\partial S}\left( -i\frac{\partial }{\partial
x^{5}}f_{5}\right) +\frac{\partial }{\partial V}\left( -i\frac{\partial }{%
\partial x^{6}}f_{5}\right)   \notag \\
&&+\lim_{\epsilon \rightarrow 0}\frac{\epsilon }{2!}\left\{ 
\begin{array}{c}
\left( -i\frac{\partial }{\partial x^{5}}\right) \left( -i\frac{\partial }{%
\partial x^{5}}f_{5}\right) \frac{\partial ^{2}}{\partial S^{2}}+2\left( -i%
\frac{\partial }{\partial x^{5}}\right) \left( -i\frac{\partial }{\partial
x^{6}}f_{5}\right) \frac{\partial ^{2}}{\partial S\partial V} \\ 
+\left( -i\frac{\partial }{\partial x^{6}}\right) \left( -i\frac{\partial }{%
\partial x^{6}}f_{5}\right) \frac{\partial ^{2}}{\partial V^{2}}+\frac{%
\partial ^{2}f_{5}}{\left( \partial x^{5}\right) ^{2}}\left( dx^{5}\right)
^{2}+\frac{\partial ^{2}f_{5}}{\left( \partial x^{6}\right) ^{2}}\left(
dx^{6}\right) ^{2} \\ 
+\left( -i\frac{\partial }{\partial x^{5}}f_{5}\right) dx^{5}\frac{\partial
^{2}}{\partial x^{5}\partial S}+\left( -i\frac{\partial }{\partial x^{6}}%
f_{5}\right) dx^{6}\frac{\partial ^{2}}{\partial x^{6}\partial V} \\ 
+2\frac{\partial ^{2}f_{5}}{\partial x^{5}\partial x^{6}}dx^{5}dx^{6}%
\end{array}%
\right\} \}.  \notag
\end{eqnarray}%
Squaring the two constrained equations, we have%
\begin{eqnarray}
\left( \frac{\partial f_{5}}{\partial S}dS+\frac{\partial f_{5}}{\partial
x^{5}}dx^{5}\right) ^{2} &=&\left( \frac{\partial f_{5}}{\partial S}%
dS\right) ^{2} \\
\left( \frac{\partial f_{5}}{\partial S}dS\right) ^{2}+2\frac{\partial f_{5}%
}{\partial S}dS\frac{\partial f_{5}}{\partial x^{5}}dx^{5}+\left( \frac{%
\partial f_{5}}{\partial x^{5}}dx^{5}\right) ^{2} &=&\left( \frac{\partial
f_{5}}{\partial S}dS\right) ^{2}  \notag \\
2\left( \frac{\partial f_{5}}{\partial S}\cdot \frac{\partial f_{5}}{%
\partial x^{5}}\right) dx^{5}dS+\left( \frac{\partial f_{5}}{\partial x^{5}}%
dx^{5}\right) ^{2} &=&0.  \notag
\end{eqnarray}%
Since $\left( \frac{\partial f_{5}}{\partial S}\cdot \frac{\partial f_{5}}{%
\partial x^{5}}\right) \neq 0$, and using $\left( \ref{cons 4}\right) $%
\begin{eqnarray}
\left( dx^{5}\right) ^{2} &=&-2\left( \frac{\partial f_{5}}{\partial S}\cdot 
\frac{\partial f_{5}}{\partial x^{5}}\right) dx^{5}dS  \notag \\
\left( dx^{5}\right) ^{2} &=&-2\left( \frac{\partial f_{5}}{\partial x^{5}}%
\right) ^{-2}\left( \frac{\partial f_{5}}{\partial S}\cdot \frac{\partial
f_{5}}{\partial x^{5}}\right) dx^{5}dS  \notag \\
\left( dx^{5}\right) ^{2} &=&-2\left( \frac{\partial x^{5}}{\partial f_{5}}%
\right) ^{2}\left( \frac{\partial f_{5}}{\partial S}\cdot \frac{\partial
f_{5}}{\partial x^{5}}\right) dx^{5}dS  \notag \\
\left( dx^{5}\right) ^{2} &=&-2\frac{\partial x^{5}}{\partial S}dx^{5}dS=0
\end{eqnarray}%
The truncated Taylor series is simplified further and yields%
\begin{eqnarray}
df_{5}\left( \mathbf{z}^{\mu }\right)  &=&\lim_{\epsilon \rightarrow 0}\frac{%
\Pi }{\epsilon }\left\{ 
\begin{array}{c}
f_{5}(t+\epsilon ,S(t+\epsilon ),V(t+\epsilon ),x^{5}(t+\epsilon
),x^{6}(t+\epsilon )) \\ 
-f_{5}(t,S(t),V(t),x^{5}(t),x^{6}(t))%
\end{array}%
\right\} dt  \notag \\
df_{5}\left( \mathbf{z}^{\mu }\right)  &=&\Pi \{\frac{\partial f_{5}}{%
\partial t}dt+\frac{\partial }{\partial S}\left( -i\frac{\partial }{\partial
x^{5}}f_{5}\right) +\frac{\partial }{\partial V}\left( -i\frac{\partial }{%
\partial x^{6}}f_{5}\right)   \label{Extra dimensional MG equation 3} \\
&&+\lim_{\epsilon \rightarrow 0}\frac{\epsilon }{2!}\left[ 
\begin{array}{c}
\left( -i\frac{\partial }{\partial x^{5}}\right) \left( -i\frac{\partial }{%
\partial x^{5}}f_{5}\right) ^{2}\frac{\partial ^{2}}{\partial S^{2}}+2\left(
-i\frac{\partial }{\partial x^{5}}\right) \left( -i\frac{\partial }{\partial
x^{6}}f_{5}\right) \frac{\partial ^{2}}{\partial S\partial V} \\ 
+\left( -i\frac{\partial }{\partial x^{6}}\right) \left( -i\frac{\partial }{%
\partial x^{6}}f_{5}\right) ^{2}\frac{\partial ^{2}}{\partial V^{2}}+\left(
-i\frac{\partial }{\partial x^{5}}f_{5}\right) dx^{5}\frac{\partial ^{2}}{%
\partial x^{5}\partial S} \\ 
+\left( -i\frac{\partial }{\partial x^{6}}f_{5}\right) dx^{6}\frac{\partial
^{2}}{\partial x^{6}\partial V}+2\frac{\partial ^{2}f_{5}}{\partial
x^{5}\partial x^{6}}dx^{5}dx^{6}%
\end{array}%
\right] \}.  \notag
\end{eqnarray}%
The derived series expansion $\left( \ref{Extra dimensional MG equation 3}%
\right) $, written in terms of superprojection operators is in fact
equivalent to the series expansion used to derive the Merton-Garman equation 
$\left( \ref{Stochastic GM equation}\right) $ with stochastic volatility.
Using equations $\left( \ref{higher dimen oper 1}\right) $ and $\left( \ref%
{higher dimen oper 2}\right) $ to project the stochastic behavior onto the
ordinary space, the series yields%
\begin{eqnarray}
df_{5}\left( \mathbf{z}^{\mu }\right)  &=&\lim_{\epsilon \rightarrow 0}\frac{%
\Pi }{\epsilon }\left\{ 
\begin{array}{c}
f_{5}(t+\epsilon ,S(t+\epsilon ),V(t+\epsilon ),x^{5}(t+\epsilon
),x^{6}(t+\epsilon )) \\ 
-f_{5}(t,S(t),V(t),x^{5}(t),x^{6}(t))%
\end{array}%
\right\} dt  \notag \\
df_{5}\left( \mathbf{z}^{\mu }\right)  &=&\dsum\limits_{n=0}^{\infty
}\dsum\limits_{m=0}^{\infty }\{\left[ 
\begin{array}{c}
\frac{\partial f_{5}}{\partial t}dt+n\frac{\partial f_{5}}{\partial S}\left(
\phi S+\sigma SR\right) dt \\ 
+m\frac{\partial f_{5}}{\partial V}\left( \mu V+\xi VQ\right) dt%
\end{array}%
\right]   \label{Extra dimensional MG equation 4} \\
&&+\lim_{\epsilon \rightarrow 0}\frac{\epsilon }{2!}\left[ 
\begin{array}{c}
n^{2}\frac{\partial ^{2}}{\partial S^{2}}f_{5}\left[ \left( \phi S+\sigma
SR\right) dt\right] ^{2} \\ 
+2nm\frac{\partial ^{2}}{\partial S\partial V}f_{5}\left[ \left( \phi
S+\sigma SR\right) dt\right] \left[ \left( \mu V+\xi VQ\right) dt\right]  \\ 
+m^{2}\frac{\partial ^{2}}{\partial V^{2}}f_{5}\left[ \left( \mu V+\xi
VQ\right) dt\right] ^{2} \\ 
+n\frac{\partial ^{2}}{\partial x^{5}\partial S}f_{5}\left[ \left( \phi
S+\sigma SR\right) dt\right] dx^{5}+2\frac{\partial ^{2}}{\partial
x^{5}\partial x^{6}}f_{5}dx^{5}dx^{6} \\ 
+m\frac{\partial ^{2}}{\partial x^{6}\partial V}f_{5}\left[ \left( \mu V+\xi
VQ\right) dt\right] dx^{6}+%
\end{array}%
\right] \}.  \notag
\end{eqnarray}%
Taking the limit $\epsilon \rightarrow 0,$ dividing both sides of equation $%
\left( \ref{Extra dimensional MG equation 4}\right) $ by $dt$ and separating
the equation into the deterministic and stochastic parts, the series reduces 
\begin{equation}
\frac{df_{5}}{dt}=\frac{\partial f_{5}}{\partial t}+\dsum\limits_{n=0}^{%
\infty }\dsum\limits_{m=0}^{\infty }\left\{ 
\begin{array}{c}
n\phi S\frac{\partial f_{5}}{\partial S}+m\mu V\frac{\partial f_{5}}{%
\partial V} \\ 
+\frac{1}{2!}\left[ 
\begin{array}{c}
n^{2}\sigma ^{2}S^{2}\frac{\partial ^{2}f_{5}}{\partial S^{2}}+m^{2}\xi
^{2}V^{2}\frac{\partial ^{2}f_{5}}{\partial V^{2}} \\ 
+2nm\sigma ^{3}S\xi \rho \frac{\partial ^{2}f_{5}}{\partial S\partial V}%
\end{array}%
\right]  \\ 
+n\sigma S\frac{\partial f_{5}}{\partial S}R+m\xi V\frac{\partial f_{5}}{%
\partial V}Q%
\end{array}%
\right\} .  \label{Generalized GM equ}
\end{equation}%
Let's simplify equation $\left( \ref{Generalized GM equ}\right) $ by
defining the following equation%
\begin{equation}
\frac{df_{5}}{dt}=f_{5}\omega _{nm}+f_{5}\alpha _{n}R+f_{5}\delta _{m}Q,
\label{simplified Generalized GM equ}
\end{equation}%
where 
\begin{eqnarray}
\omega _{nm} &=&\frac{\partial }{\partial t}+\dsum\limits_{n=0}^{\infty
}\dsum\limits_{m=0}^{\infty }\left\{ 
\begin{array}{c}
n\phi S\frac{\partial }{\partial S}+m\mu V\frac{\partial }{\partial V} \\ 
+\frac{1}{2!}\left[ 
\begin{array}{c}
n^{2}\sigma ^{2}S^{2}\frac{\partial ^{2}}{\partial S^{2}}+m^{2}\xi ^{2}V^{2}%
\frac{\partial ^{2}}{\partial V^{2}} \\ 
+2nm\sigma ^{3}S\xi \rho \frac{\partial ^{2}}{\partial S\partial V}%
\end{array}%
\right] 
\end{array}%
\right\} ,  \label{sub 1} \\
\alpha _{n} &=&n\sigma S\frac{\partial }{\partial S},  \label{sub 2} \\
\delta _{m} &=&m\xi V\frac{\partial }{\partial V}.  \label{sub 3}
\end{eqnarray}%
We consider a following self-replicating higher-dimensional portfolio%
\begin{equation}
\pi _{5}(t)=\theta _{1}f_{5}(t)+\theta _{2}S(t),
\end{equation}%
where $\theta _{1}$ and $\theta _{2}$ are the amounts of option $f_{5}$ and
stock $S,$ respectively. The total time derivative of our portfolio is given%
\begin{equation}
\frac{d\pi _{5}}{dt}=\theta _{1}\frac{df_{5}}{dt}+\theta _{2}\frac{dS}{dt}.
\end{equation}%
Using equations $\left( \ref{simplified Generalized GM equ}\right) ,$ the
total time-derivative for out portfolio yields%
\begin{eqnarray}
\frac{d\pi _{5}}{dt} &=&\theta _{1}\left( f_{5}\omega _{nm}+f_{5}\alpha
_{n}R+f_{5}\delta _{m}Q\right) +\theta _{2}\left( \phi S+\sigma SR\right)  
\notag \\
\frac{d\pi _{5}}{dt} &=&\omega _{nm}\theta _{1}f_{5}+\phi \theta
_{2}S+\left( \alpha _{n}R+\delta _{m}Q\right) \theta _{1}f_{5}+\theta
_{2}\sigma SR.  \label{evo-port}
\end{eqnarray}%
We choose our portfolio in such a way that it satisfies the following
hedging equations%
\begin{eqnarray}
\alpha _{n}R\theta _{1}f_{5}+\theta _{2}\sigma SR &=&0,
\label{matrix relation 1} \\
\delta _{m}Q\theta _{1}f_{5} &=&0.  \label{matrix relation 2}
\end{eqnarray}%
The random quantities $R(t)$ and $Q(t)$ can be automatically eliminated from 
$\left( \ref{evo-port}\right) $ with the aid of equations $\left( \ref%
{matrix relation 1}\right) $ and $\left( \ref{matrix relation 2}\right) .$
Thus, written in matrix notation, equations $\left( \ref{matrix relation 1}%
\right) $ and $\left( \ref{matrix relation 2}\right) $ yields the following
matrix equation 
\begin{equation}
\left[ 
\begin{array}{cc}
\alpha _{n} & \sigma  \\ 
\delta _{m} & 0%
\end{array}%
\right] \left[ 
\begin{array}{c}
\theta _{1}f_{5} \\ 
\theta _{2}S%
\end{array}%
\right] =0.  \label{matrix relation 3}
\end{equation}%
Subsequently, the time-derivative of the portfolio $\left( \ref{evo-port}%
\right) $ reduces to a pure deterministic equation%
\begin{equation}
\frac{d\pi _{5}}{dt}=\omega _{nm}\theta _{1}f_{5}+\phi \theta _{2}S.
\label{deterministic dpi/dt}
\end{equation}%
The absence of arbitrage forces the time-derivative of the portfolio to be
directly proportional to the risk-neutral rate $r$, hence 
\begin{equation}
\frac{d\pi _{5}}{dt}=r\pi _{5}.  \label{risk-free growth}
\end{equation}%
Solving equation $\left( \ref{risk-free growth}\right) $ for the
higher-dimensional portfolio, we have%
\begin{eqnarray}
\frac{d\pi _{5}}{\pi _{5}} &=&rdt  \notag \\
\ln \left\vert \pi _{5}\right\vert  &=&\int rdt=rt+C  \notag \\
\pi _{5}(t) &=&\pi _{0}\exp (rt),
\end{eqnarray}%
where $C$ is a integration constant and $\pi _{0}=\exp (C)$. Continue with
the derivation of the Merton-Garman equation, we note that equality of
equations $\left( \ref{deterministic dpi/dt}\right) $ and $\left( \ref%
{risk-free growth}\right) $ yields%
\begin{equation}
\left( \omega _{nm}-r\right) \theta _{1}f_{5}+\left( \phi -r\right) \theta
_{2}S=0,
\end{equation}%
and in matrix notation 
\begin{equation}
\left[ 
\begin{array}{cc}
\omega _{nm}-r & \phi -r%
\end{array}%
\right] \left[ 
\begin{array}{c}
\theta _{1}f_{5} \\ 
\theta _{2}S%
\end{array}%
\right] =0.  \label{matrix relation 4}
\end{equation}%
Equation $\left( \ref{matrix relation 4}\right) $ can only be satisfied if 
\begin{equation}
\left[ 
\begin{array}{c}
\lambda _{1}(t) \\ 
\lambda _{2}(t)%
\end{array}%
\right] \left[ 
\begin{array}{cc}
\omega _{nm}-r & \phi -r%
\end{array}%
\right] =\left[ 
\begin{array}{cc}
\alpha _{n} & \sigma  \\ 
\delta _{m} & 0%
\end{array}%
\right] .  \label{eigen valu equa}
\end{equation}%
Solving equation $\left( \ref{eigen valu equa}\right) $ for $\left[ 
\begin{array}{cc}
\omega _{nm}-r & \phi -r%
\end{array}%
\right] $, we have 
\begin{equation}
\left[ 
\begin{array}{cc}
\omega _{nm}-r & \phi -r%
\end{array}%
\right] =\left[ 
\begin{array}{cc}
\lambda _{1}^{\ast }(t) & \lambda _{2}^{\ast }(t)%
\end{array}%
\right] \left[ 
\begin{array}{cc}
\alpha _{n} & \sigma  \\ 
\delta _{m} & 0%
\end{array}%
\right] ,  \label{eigen valu equa 2}
\end{equation}%
where we have $\lambda _{1}^{\ast }(t)\lambda _{1}(t)+\lambda _{2}^{\ast
}(t)\lambda _{2}(t)=\lambda _{1}^{2}(t)+\lambda _{2}^{2}(t)=1.$ In component
forms, the matrix equation $\left( \ref{eigen valu equa 2}\right) $ yields%
\begin{eqnarray}
\omega _{nm}-r &=&\lambda _{1}(t)\alpha _{n}+\lambda _{2}(t)\delta _{m}
\label{component 1} \\
\phi -r &=&\lambda _{1}(t)\sigma .  \label{component 2}
\end{eqnarray}%
Solving equations $\left( \ref{component 1}\right) $ and $\left( \ref%
{component 2}\right) $ simultaneously, we have 
\begin{equation}
\phi -r=\lambda _{1}\sigma =\left( \frac{\omega _{nm}-r}{\alpha _{n}}-\frac{%
\lambda _{2}(t)\delta _{m}}{\alpha _{n}}\right) \sigma ,  \label{component 3}
\end{equation}%
where $\lambda _{1}(t)=\frac{\omega _{nm}-r}{\alpha _{n}}-\frac{\lambda
_{2}(t)\delta _{m}}{\alpha _{n}}.$ Solving for $\omega _{nm}-r,$ we then
have 
\begin{equation}
\omega _{nm}-r=\left( \frac{\phi -r}{\sigma }\right) \alpha _{n}+\lambda
_{2}(t)\delta _{m}.
\end{equation}%
We define a $5D$ Merton-Garman operator $\Gamma _{5}$ by 
\begin{equation}
\Gamma _{nm}=\left( \omega _{nm}-r\right) -\left( \frac{\phi -r}{\sigma }%
\right) \alpha _{n}-\lambda _{2}(t)\delta _{m}.
\end{equation}%
In term of $5D$ Merton-Garman operator $\Gamma _{5},$ the generalized $5D$
Merton-Garman equation takes the compact form%
\begin{equation}
\Gamma _{\left( n,m\right) }f_{5}=0.  \label{5D MG equation}
\end{equation}%
The ordinary $3D$ Merton-Garman equation can now be obtained from equation $%
\left( \ref{5D MG equation}\right) $ for $n=m=1$%
\begin{eqnarray}
\Gamma _{\left( 1,1\right) }f_{5} &=&0  \label{3D Merton-Garman} \\
\left[ \left( \omega _{11}-r\right) -\left( \frac{\phi -r}{\sigma }\right)
\alpha _{1}-\lambda _{2}\delta _{1}\right] f_{5} &=&0.  \notag
\end{eqnarray}%
To demonstrate the equivalency between equation $\left( \ref{3D
Merton-Garman}\right) $ and equation $\left( \ref{Stochastic GM equation 3}%
\right) ,$ we use equations $\left( \ref{sub 1}\right) ,$ $\left( \ref{sub 2}%
\right) ,$ $\left( \ref{sub 3}\right) $ and $\left( \ref{Hull}\right) $ in
the dimensional reduction of the higher-dimensional Merton-Garman equation $%
\left( \ref{5D MG equation}\right) ,$which yields 
\begin{eqnarray}
0 &=&\Gamma _{\left( 1,1\right) }f_{5}=\left[ \left( \omega _{11}-r\right)
-\left( \frac{\phi -r}{\sigma }\right) \alpha _{1}-\lambda _{2}\delta _{1}%
\right] f_{5} \\
0 &=&\frac{\partial }{\partial t}f_{5}+\frac{1}{2!}\left\{ \sigma ^{2}S^{2}%
\frac{\partial ^{2}}{\partial S^{2}}f_{5}+\xi ^{2}V^{2}\frac{\partial ^{2}}{%
\partial V^{2}}f_{5}+2\sigma ^{3}S\xi \rho \frac{\partial ^{2}}{\partial
S\partial V}f_{5}\right\}   \notag \\
&&-rf_{5}+rS\frac{\partial }{\partial S}f_{5}+\overline{\mu }\sigma ^{2}%
\frac{\partial }{\partial S}f_{5}.  \notag
\end{eqnarray}%
Since the stochastic information had already been mapped onto the ordinary
space, we can then safely replace $f_{5}$ by $f,$ hence%
\begin{eqnarray}
0 &=&\frac{\partial }{\partial t}f+\frac{1}{2!}\left\{ \sigma ^{2}S^{2}\frac{%
\partial ^{2}}{\partial S^{2}}f+\xi ^{2}V^{2}\frac{\partial ^{2}}{\partial
V^{2}}f+2\sigma ^{3}S\xi \rho \frac{\partial ^{2}}{\partial S\partial V}%
f\right\}   \label{Reduced Merton-Garman equation} \\
&&-rf+rS\frac{\partial }{\partial S}f+\overline{\mu }\sigma ^{2}\frac{%
\partial }{\partial S}f.  \notag
\end{eqnarray}%
By inspection, equation $\left( \ref{Reduced Merton-Garman equation}\right) $
is identical to equation $\left( \ref{Stochastic GM equation 3}\right) $
from section 2. Thus, we have shown that the generalized Black-Scholes
equation or the Merton-Garman equation can be obtained my the
higher-dimensional approach along with two constraints.

\section{Summary}

We showed that the celebrated Black-Scholes equation and its generalized
version Merton-Garman equation can be obtained from an enlarged manifold.
The Merton-Garman equation was shown to be an equation of the first excited
state i.e. $n=m=1.$ In fact, there are an infinite number of the
Merton-Garman-like equations contained in our superspace. These equations
could be called excited states living in superspace. These excited states
only manifest their presence in our ordinary space when acted upon by the
superprojection operators $P_{5}$ and $P_{6}.$ For the ground state, i.e. $%
n=m=0,$ nothing is projected from superspace. In other words, the stochastic
or random nature of the variables is confined in the extra dimensional
subspace of superspace. In general, in order to extract forecasting aspects
or predictive power of a financial theory, we need to recast the
Merton-Garman equation into its quantum mechanical Schrodinger form. In
quantum mechanical form, manipulation of pertinent information such as
(predictive power, hedging, and arbitrage...) associated with a financial
theory can be executed via the potential functions, i.e. the Hamiltonians or
Lagrangians. Furthermore, the potential functions can also be influenced by
the nature of the extra dimensions. The effects of the potential functions
by the extra dimensions along with the Schrodinger interpretation of the
classical theory, will then provide financial theorists or model builders
with alternate research avenues and a larger theoretic framework in which
financial theories are obtained.

\pagebreak


\section{References}

\begin{thebibliography}{99}
\bibitem{Black-Scholes} F. Black and M. Scholes. "The Pricing of Options and
Corporate Liabilities." Journal of Political Economy 81 (May 1973), 637-659

\bibitem{Merton} R. C. Merton. "The Theory of Rational Option Pricing." Bell
Journal of Economics and Management Science 4 (Spring 1973), 141-83

\bibitem{Garman} M. Garman. "A General Theory of Asset Valuation under
diffusion State Processes." Working Paper No 50, University of California,
Berkeley, 1976.

\bibitem{Bouchaud} J. -P. Bouchaud and D. Sornette. "The Black-Scholes
Option Pricing Problem in Mathematical Finance." Journal de Physique I 4.
863-81 (1994).

\bibitem{Baaquie} B. E. J. Phys, I (France) 7, 1733-1753, (1997)

\bibitem{Linetsky} V Litnetsky. "The Path Integral Approach to Financial
Modelling and Options Pricing." Computational Economics 11, 129. 1998.

\bibitem{Truong} M. Truong, Physical Review D74 (2006).

\bibitem{Hull} J. C. Hull and A. White. "The Pricing of Options on Assets
with Stochastic Volatilities." The Journal of Finance, Vol XLII, No. 2 (June
1987),281-299.

\bibitem{Kaluza} T. Kaluza. Sitzungdber. Berl. Akad. (1921) 966.

\bibitem{Klein} O. Z. Klein. Phyz. 37 (1926) 895.
\end{thebibliography}
\end{document}